\documentclass{iaus}
\usepackage{graphicx}

\title[Hyperfast neutron stars] 
{On the origin of hyperfast neutron stars}

\author[Gvaramadze et al.]   
{V.V. Gvaramadze$^1$%
  , A. Gualandris$^{2}$ \break \and S. Portegies Zwart$^3$}

\affiliation{$^1$Sternberg Astronomical Institute, Moscow State
University, Universitetskij Pr. 13, Moscow 119992, Russia
\break email: vgvaram@mx.iki.rssi.ru\\[\affilskip]
$^2$Center for Computational Relativity and Gravitation, Rochester
Institute of Technology, 85 Lomb Memorial Drive, Rochester, NY
14623, USA \break email: alessiag@astro.rit.edu\\[\affilskip]
$^3$ Astronomical Institute `Anton Pannekoek' and Section
Computational Science, University of Amsterdam, Kruislaan 403, 1098
SJ, Amsterdam, the Netherlands \break email: spz@science.uva.nl}

\pubyear{2008}
\volume{246}  
\pagerange{119--126}
\date{?? and in revised form ??}
 \jname{Dynamical Evolution of Dense Stellar
Systems}
\editors{ E. Vesperini, M. Giersz, \& A. Sills, eds.}
\begin{document}

\maketitle

\begin{abstract}
We propose an explanation for the origin of hyperfast neutron stars
(e.g. PSR B1508+55, PSR B2224+65, RX J0822$-$4300) based on the
hypothesis that they could be the remnants of a {\it symmetric}
supernova explosion of a high-velocity massive star (or its helium
core) which attained its peculiar velocity (similar to that of the
neutron star) in the course of a strong three- or four-body
dynamical encounter in the core of a young massive star cluster.
This hypothesis implies that the dense cores of star clusters
(located either in the Galactic disk or near the Galactic centre)
could also produce the so-called hypervelocity stars -- the ordinary
stars moving with a speed of $\sim 1\,000 \, {\rm km} \, {\rm
s}^{-1}$. 
\end{abstract}

\firstsection 
\section{Introduction}

Recent proper motion and parallax measurements for the pulsar PSR
B1508+55 (Chatterjee et al. \cite{cha05}) gave the first example of
a high velocity ($1\,083_{-90} ^{+103} \, {\rm km} \, {\rm s}^{-1}$)
{\it directly} measured for a neutron star (NS). A possible way to
account for extremely high velocities of NSs\footnote{Other possible
examples of hyperfast NSs are PSR B2224+65 (Chatterjee \& Cordes
\cite{cha04}) and RX J0822-4300 (Hui \& Becker \cite{hui06}).} is to
assume that they are due to a natal kick or a post-natal
acceleration (Chatterjee et al. 2005). In this paper, we propose an
alternative explanation for the origin of hyperfast NSs (cf.
Gvaramadze \cite{gva07}) based on the hypothesis that they could be
the remnants of {\it symmetric} supernova (SN) explosions of
hypervelocity stars [HVSs; the ordinary stars moving with extremely
high ($\sim 1\,000 \, {\rm km} \, {\rm s}^{-1}$) peculiar
velocities; e.g. Brown et al. \cite{bro05}]. A strong argument in
support of this hypothesis comes from the fact that the mass of one
of the HVSs is $\geq 8 \, M_{\odot}$ (Edelmann et al. \cite{ede05})
so that this star ends its evolution in a type II SN leading to the
production of a hyperfast NS.

\section{Hypervelocity stars and young massive star clusters}

It is believed that the origin of HVSs could be connected to
scattering processes involving the supermassive black hole (BH) in
the Galactic centre (Hills \cite{hil88}; Yu \& Tremaine \cite{yu03};
Gualandris et al. \cite{gua05}). It is therefore possible that the
progenitors of some hyperfast NSs were also ejected from the
Galactic centre. The proper motion and parallax measured for PSR
B1508+55, however, indicate that this NS was born in the Galactic
disk (Chatterjee et al. \cite{cha05}). The kinematic characteristics
of some high-velocity early B stars also suggest that these objects
originated in the disk (e.g. Ramspeck et al. \cite{ram01}). We
consider the possibility that the HVSs (including the progenitors of
hyperfast NSs) could be ejected not only from the Galactic centre
but also from the cores of young ($<10^7$ yr) massive ($\sim 10^4 -
10^5 \, M_{\odot}$) star clusters (YMSCs), located either in the
Galactic disk or near the Galactic centre (cf. Gualandris \&
Portegies Zwart \cite{gua07}).

\section{Origin of hyperfast neutron stars}

To check the hypothesis that the hyperfast NSs could be the
descendants of HVSs which were ejected from the cores of YMSCs, we
calculated (see Gvaramadze et al. \cite{gva08}) the maximum possible
ejection speed produced by dynamical processes involving close
encounters between: {\it i}) two hard (Heggie \cite{heg75}) massive
binaries (e.g. Leonard \cite{leo91}), {\it ii}) a hard binary and an
intermediate-mass ($\sim 100-1\,000 \, M_{\odot}$) BH (e.g.
Portegies Zwart \& McMillan \cite{por02}), and {\it iii}) a single
star and a hard binary intermediate-mass BH (e.g. G\"{u}rkan et al.
\cite{gur06}). We find that main-sequence O-type stars cannot be
ejected from YMSCs with peculiar velocities high enough to explain
the origin of hyperfast NSs, but lower mass main-sequence stars or
the stripped helium cores of massive stars could be accelerated to
hypervelocities. We find also that the dynamical processes in the
cores of YMSCs can produce stars moving with velocities of $\sim
200-400 \, {\rm km} \, {\rm s}^{-1}$ which therefore contribute to
the origin of high-velocity NSs as well as to the origin of the
bound population of halo stars (Ramspeck et al. \cite{ram01}; Brown
et al. \cite{bro07}).

\begin{acknowledgments}
V.\,V.\,G. acknowledges the International Astronomical Union and the
Russian Foundation for Basic Research for travel grants. A.\,G. is
supported by grant NNX07AH15G from NASA. S.\,P.\,Z. acknowledge
support from the Netherlands Organization for Scientific Research
(NWO under grant No. 635.000.001 and 643.200.503) and the
Netherlands Research School for Astronomy (NOVA).
\end{acknowledgments}

\end{document}